\documentclass[12pt, prd, showpacs]{revtex4}
\usepackage{amssymb}
\usepackage{amsmath}
\usepackage{graphicx}

\input{tcilatex}

\begin{document}

\title{Strategies of motion under the black hole horizon}
\author{A. \ V. Toporensky}
\affiliation{Sternberg Astronomical Institute, Lomonosov Moscow State University }
\affiliation{Faculty of Physics, Higher School of Economics, Moscow, Russia\\
Kazan Federal University, Kremlevskaya 18, Kazan 420008, Russia}
\email{atopor@rambler.ru}
\author{O. B. Zaslavskii}
\affiliation{Department of Physics and Technology, Kharkov V.N. Karazin National
University, 4 Svoboda Square, Kharkov 61022, Ukraine}
\affiliation{Kazan Federal University, Kremlevskaya 18, Kazan 420008 Russia}
\email{zaslav@ukr.net}

\begin{abstract}
In this methodological paper we consider two problems an astronaut faces
with under the black hole horizon in the Schwarzschild metric. 1) How to
maximize the survival proper time. 2) How to make a visible part of the
outer Universe as large as possible before hitting the singularity. Our
consideration essentially uses the concept of peculiar velocities based on
the "river model". Let an astronaut cross the horizon from the outside. We
reproduce from the first principles the known result that point 1) requires
that an astronaut turn off the engine near the horizon and follow the path
with the momentum equal to zero. We also show that point 2) requires
maximizing the peculiar velocity of the observer. Both goals 1) and 2)
require, in general, different strategies inconsistent with each other that
coincide at the horizon only. The concept of peculiar velocities introduced
in a direct analogy with cosmology, and its application for the problems
studied in the present paper can be used in advanced general relativity
courses.
\end{abstract}

\maketitle

\section{Introduction}

"Elementary" metrics of black holes such as the Schwarzschild or
Reissner-Nordstr\"{o}m ones are studied very thoroughly and described in
many textbooks. Nonetheless, time to time new and, sometimes, quite
unexpected aspects of them appear again and again. The aim of the present
article is to draw attention to the special class of trajectories situated
inside the event horizon of a nonextremal black hole. These trajectories
correspond to $\varepsilon =0$ where the integral of motion $\varepsilon $
has inside a horizon the meaning of momentum, not energy. And, among them
the most interesting is their subclass when the trajectory strongly bends to
the horizon. The results are valid for a quite generic metric but, for
definiteness, we concentrate mainly on the Schwarzschild one. There is a
number of different issues in this context which, however, turned out to be
interrelated due to this class of trajectories. In our previous work \cite%
{zero}, we considered high-energy collisions of a particle that follows such
a trajectory with ambient ones. Meanwhile, there are also other interesting
properties of the same class of trajectories discussed in the present work.
Namely, we consider the following questions: (i) the maximization of time
before reaching the singularity, (ii) possibility for an observer to see
during a finite proper time all future of Universe (this issue was discussed
in \cite{kras}, \cite{gpu} but account of the trajectories under discussion
adds new options not noticed there). Both questions usually attract great
attention of a student audience while General Relativity studying. The
concept of "river of space" itself, being a counterpart of the famous
concept of "expanding space" in cosmology, as well as the definition of
peculiar velocities introduced in a direct analogy of the corresponding
definition in cosmology can be used in advanced General Relativity courses.
They show in particular that some intuitively clear concepts in GR represent
not the physical situation itself, but mostly the properties of a coordinate
system used to describe this picture. Using appropriate coordinate system
better suited for a physical problem considered is a necessary skill for
scientific work in GR. We hope that the present paper where we study the
spherically symmetric static metric in three different frames (one
stationary and two different synchronous frames) depending on a particular
question considered, would help in achieving this goal.

Throughout the paper, we use geometric units in which fundamental constants $%
G=c=1$.

\section{Basic equations}

We consider the metric%
\begin{equation}
ds^{2}=-fdt^{2}+\frac{dr^{2}}{f}+r^{2}d\omega ^{2}\text{,}  \label{met}
\end{equation}%
where $d\omega ^{2}=(d\theta ^{2}+\sin ^{2}\theta d\phi ^{2}).$

We suppose that the metric has the event horizon at $r=r_{+}$, so $%
f(r_{+})=0 $. For the Schwarzschild metric, $f=1-\frac{r_{+}}{r}$, where $%
r_{+}=2M$ is the horizon radius, $M$ being the black hole mass. The most
part of results applies also to generic $f(r)$ with one root. Near the event
horizon,

\begin{equation}
f\approx \kappa (r-r_{+})\text{,}  \label{kar}
\end{equation}%
where $\kappa =\frac{f^{\prime }(r_{+})}{2}$ is the surface gravity. In the
Schwarzschild case $\kappa =\frac{1}{2r_{+}}$.

We are mainly concerned with the interior of a black hole. Then, it is
instructive to make substitutions (see \cite{nov61} and \cite{fn}, page 25) 
\begin{equation}
r=-T,t=y,f=-\tilde{f},\text{ }\tilde{f}\geq 0,  \label{subst}
\end{equation}%
where $-r_{+}\leq T\leq 0$. The coordinate $y$ plays the role of a radial
coordinate inside the horizon while $T$ has the meaning of time.

Then,%
\begin{equation}
ds^{2}=-\frac{dT^{2}}{\tilde{f}}+\tilde{f}dy^{2}+T^{2}d\omega ^{2}\text{.}
\label{min}
\end{equation}

For the Schwarzschild metric $\tilde{f}=\frac{r_{+}}{r}-1=-1-\frac{r_{+}}{T}%
, $ where $-r_{+}<T\leq 0$.

The metric under discussion looks differently in different regions of
space-time. This is a particular manifestation of the fact that in
spherically symmetric space-times there exist so-called $R$ and $T$ regions.
Namely, let us denote $\rho ^{2}$ the coefficient at the angular part. Then,
if the vector normal to the surface $\rho =const$ is space-like, the region
is called the $R$ one. If such a normal vector is time-like, it corresponds
to the $T$ region. Such a classification was suggested in \cite{rt}. The
full space-time diagram of the Schwarazschild eternal black hole includes
two $R$ regions $R1$ and $R2$ and two $T$ regions - expanding and
contracting ones. This description can be found in many textbooks and
monographs. See, e.g. \cite{fn}. Region $R1$ corresponds to our world while $%
R2$ represents the "mirror world". The second $T$ region and $R2$ region
exist only in a mathematical picture which includes black and white eternal
hole, and they are absent in a realistic black hole space-times obtained as
a result of gravitational collapse.

In the metric (\ref{met}), (\ref{min}) there is an integral of motion%
\begin{equation}
\varepsilon =-u_{\mu }\xi ^{\mu }\text{,}  \label{e}
\end{equation}%
where $\xi ^{\mu }$ is the corresponding Killing vector responsible for
translations along the $y$ axis, $u^{\mu }=\frac{dx^{\mu }}{d\tau }$ being
the four-velocity, $\tau $ the proper time. Under the horizon, $\varepsilon $
has the meaning of the radial momentum, 
\begin{equation}
\varepsilon =-u_{y}.  \label{eu}
\end{equation}%
Three equations of motion within the plane $\theta =\frac{\pi }{2}$ for a
geodesic particle read%
\begin{equation}
\dot{T}=Z\text{,}  \label{tz}
\end{equation}%
\begin{equation}
\dot{y}=-\frac{\varepsilon }{\tilde{f}}\text{,}  \label{my}
\end{equation}%
\begin{equation}
m\dot{\phi}=\frac{L}{T^{2}}\text{,}  \label{L}
\end{equation}%
dot denotes differentiation with respect to the proper time $\tau $. Eq. (%
\ref{my}) follows immediately from (\ref{eu}) in the metric (\ref{min}). If
the Killing vector corresponds to rotations, in coordinates $(T,y,\phi )$ it
has the form ($0,0,1$), so similar to (\ref{eu}) we have $L=mu_{\phi }$. In
this manner, we get eq. (\ref{L}). Then, from the normalization condition
and $\dot{T}>0$ we obtain (\ref{tz}).

Here, $\varepsilon $ can have any sign, $\varepsilon =\pm \left\vert
\varepsilon \right\vert .$ The case $\varepsilon =0$ is also possible. From
the normalization \ condition $u_{\mu }u^{\mu }=-1$ and taking into account
the forward-in-time condition $\dot{T}>0$, we obtain in coordinates $%
(T,y,\phi )$ the four-velocity of a particle moving within the plane $\theta
=\frac{\pi }{2}$:%
\begin{equation}
u^{\mu }=(Z,-\frac{\varepsilon }{\tilde{f}},\frac{L}{mT^{2}})\text{,}
\label{muu}
\end{equation}%
\begin{equation}
u_{\mu }=(-\frac{Z}{\tilde{f}},-\varepsilon ,\frac{L}{m}),  \label{mul}
\end{equation}%
\begin{equation}
Z=\sqrt{\varepsilon ^{2}+\tilde{f}(1+\frac{L^{2}}{m^{2}T^{2}})}\text{.}
\label{zt}
\end{equation}%
From (\ref{my}), it follows that a proper time required for travel from the
horizon $r_{+}$ to $r_{1}<r_{+}$ is equal to%
\begin{equation}
\tau =\int_{r_{1}}^{r_{+}}\frac{dr}{Z}=\int_{-r_{+}}^{T_{1}}\frac{dT}{\sqrt{%
\varepsilon ^{2}+\tilde{f}(1+\frac{L^{2}}{m^{2}T^{2}})}}.  \label{tau}
\end{equation}

\section{Behavior of the proper time}

After crossing the horizon of a black hole, an observer inevitably falls
into the singularity, so the low limit in the integral (\ref{tau}) $r_{1}=0$%
. One may ask, how to make the corresponding proper time as large as
possible \cite{austr}. To maximize $\tau $, we must minimize $Z$. It is seen
from (\ref{tau}) that this is achieved if $\varepsilon =0$, $L=0$. In
particular, for the Schwarzschild metric we get the famous maximal interval
of proper time equal to 
\begin{equation}
\tau _{\max }=(\pi /2)r_{+}.  \label{tmax}
\end{equation}

In the above formulas it was assumed that a particle moves freely. What
changes if it undergoes some acceleration? Then, because of the action of
the force, the momentum $\varepsilon $ is not conserved any longer. However,
one can still use definition (\ref{e}) where now $\varepsilon =\varepsilon
(T)$ depends on time. One has $u^{y}=\frac{dy}{d\tau }=\frac{u_{y}}{\tilde{f}%
}=-\frac{\varepsilon }{\tilde{f}}$ that coincides with eq. (\ref{my}). In a
similar way, one can check the validity of (\ref{L}). Then, taking into
account the normalization condition $u_{\mu }u^{\mu }=-1$, we obtain eqs. (%
\ref{tz}) - (\ref{my}) again. In other words, the equations of motion retain
the same form as for free fall but now the quantity $\varepsilon $ is not
integral of motion.

From this simple observation, an important consequence follows. It turns out
that formula (\ref{tau}) with $Z$ given by (\ref{zt}) is also valid. This
means that the main conclusion about the maximum survival time holds true.
Namely, the geodesic trajectory with $\varepsilon =0$, $L=0$ is more
"profitable" than any other one, including those even under the action of
force.

If a particle moves along the geodesic path in a static (or homogeneous)
metric, its energy (or momentum) and angular momentum are obviously
conserved. Moreover, in our context the reverse is also true. Indeed, there
are two independent integrals of motion and there are two independent
equations of motion. Therefore, the values of these integrals fix the
trajectory (up to the constant affecting the initial moment of motion). More
precise formulation consists in that the trajectory maximizing the survival
proper time is (i) geodesic, for which (ii) $\varepsilon =0$ and $L=0$. But
if a particle fell from the outer region, $\varepsilon $ had there the
meaning of energy. There, a particle cannot have $\varepsilon =0$ exactly.
It can have very small $\varepsilon >0$, provided it began its motion very
closely to the horizon. Alternatively, a particle can move with finite
nonzero $\varepsilon \sim 1$ but very nearly to the horizon it should
experience large deceleration by some engine. To decelerate a particle near
the horizon, the corresponding engine should be super-power since the
required force diverges in the horizon limit.

In a qualitative form the discussion of the survival proper time can be
found in \cite{novengl}, pages 45 - 46. Numerically, the best strategy to
maximize the proper time -- reach the geodesics $\varepsilon =0$ as soon as
possible inside the horizon, and then stay on it -- has been confirmed in 
\cite{austr}. We presented a simple derivation of this fact from the first
principles. On the other hand, we agree with \cite{austr} that a naive first
principle "explanation" based on the minimal action principle (the curve
maximizing a proper time between two fixed points is a geodesic) is not
applicable here since the black hole singularity is not a space-time point.
It seems that these ideas became some kind of folklore. The fact that the
proper time of the fall has its maximum at the geodesic (so, if a falling
object is equipped by an engine, it is better to switch it off at some
point!) is often considered as a rather counter-intuitive, since naive
approach prompts to decelerate, whereas the true best strategy requires to
"give up" at some point.

In the Section V we introduce another time variable which also can be used
to describe a free fall into a black hole, and this variable does match an
intuitively clear strategy of "never giving up".

\section{Generalization}

The material of the previous Section admits generalization which, at the
same time, reveals the underlying reason of the best strategy of maximizing
the proper survival time. Let us consider some generic metric. Let we have
some spatial hypersurface $\Sigma $ and ask the question: how to maximize
the proper time $\tau $ from a given \ initial point O to $\Sigma $? In
doing so, the hypersurface $\Sigma $ can be singular (like in the
Schwarzschild case) or regular. Different paths with a fixed initial point
intersect $\Sigma $ in different final points, so we are unable to refer
directly to the action principle. We choose another approach. Let us
introduce a synchronous frame such that $\Sigma $ is described by equation $%
\tilde{t}=\tilde{t}_{0}$, where $\tilde{t}_{0}$ is some constant,%
\begin{equation}
ds^{2}=-d\tilde{t}^{2}+\gamma _{ik}dx^{i}dx^{k}\text{,}  \label{gen}
\end{equation}%
$\gamma _{ik\text{ }}$is some spatial metric. As is known (see \cite{LL},
Chapter 11.97), such a frame can always be constructed.

We can introduce a three-dimensional velocity of a particle according to%
\begin{equation}
v^{i}=\frac{dx^{i}}{d\tilde{t}}\text{, }v^{2}=\gamma _{ik}v^{i}v^{k}\text{.}
\end{equation}%
Along the particle trajectory, $dx^{i}=v^{i}d\tilde{t}$. By substitution
into (\ref{gen}), we obtain%
\begin{equation}
ds^{2}=-d\tilde{t}^{2}(1-v^{2})
\end{equation}%
along the trajectory.

Then, the proper time $\tau $ between O and $\Sigma $ is equal to%
\begin{equation}
\tau =\int d\tilde{t}\sqrt{1-v^{2}}\text{.}
\end{equation}

Obviously, we must choose $v=0$ to maximize $\tau $. Then, an observer
remains at rest and follows the path $x^{i}=const$ which is geodesic. The
conclusion under discussion does not require the ability to integrate the
equations of motion. This is important since in general such integration is
possible only for the metrics with some symmetry properties like it happens
for metric (\ref{met}).

\section{Two synchronous frames under black hole horizon}

It is well-known and described in many textbooks, that using example of a
synchronous frame called a Lema\^{\i}tre one, enables us to build a frame
that behaves regularly on the horizon of a black hole. Meanwhile, in the
present Section we wish to stress here some important subtleties according
to which there are at least two different relevant synchronous frames under
the horizon. To make presentation self-contained, we give here brief
description of the frames we use further in the present paper (details can
be found in \cite{ham}, \cite{zero}, \cite{3}). The approach is based on the
Painlev\'{e}-Gullstrand frame for the Schwarzschild metric \cite{pain}, \cite%
{gull} or its generalization. Namely, starting from (1) one can introduce a
new time coordinate according to 
\begin{equation}
\tilde{t}=t+\int \frac{dr}{f}\sqrt{1-f}\text{,}  \label{tt}
\end{equation}

Then, the metric takes the form%
\begin{equation}
ds^{2}=-fd\tilde{t}^{2}+dr^{2}+2d\tilde{t}drv+r^2d\omega ^{2}\text{,}
\label{metgp}
\end{equation}%
where 
\begin{equation}
v=\sqrt{1-f}\text{.}  \label{vg}
\end{equation}

The relevant contravariant components of the metric are equal to 
\begin{equation}
g^{\tilde{t}\tilde{t}}=-1\text{, }g^{\tilde{t}r}=-v\text{, }g^{rr}=f\text{.}
\end{equation}

The metric can be written as%
\begin{equation}
ds^{2}=-d\tilde{t}^{2}+(vd\tilde{t}+dr)^{2}+r^{2}d\omega ^{2}\text{,}
\label{gp}
\end{equation}%
If, in addition to (\ref{tt}), one transforms also the radial coordinate
according to%
\begin{equation}
\rho =t+\int \frac{dr}{f\sqrt{1-f}}\text{,}  \label{ry}
\end{equation}%
we obtain the generalized Lema\^{\i}tre frame that turns into a standard one
for $f=1-\frac{r_{+}}{r}$ (see, e.g. \cite{LL}, Chapter 12. Sec. 102), 
\begin{equation}
ds^{2}=-d\tilde{t}^{2}+d\rho ^{2}(1-f)+r^{2}(\rho ,\tilde{t})d\omega ^{2}%
\text{.}  \label{metlem}
\end{equation}

For shortness, we will call $\tilde{t}$ the Lema\^{\i}tre time. The frame is
regular in the vicinity of the horizon, where $f=0=\tilde{f}$.

Note, however, that the frame (\ref{min}), applicable under the horizon, can
be also considered as an example of the synchronous frame, where the
corresponding synchronous time is equal to 
\begin{equation}
\hat{t}=\int \frac{dT}{\sqrt{\tilde{f}(T)}},  \label{tn}
\end{equation}%
so metric (\ref{min}) takes the form%
\begin{equation}
ds^{2}=-d\hat{t}^{2}+\tilde{f}dy^{2}+T^{2}(\hat{t})d\omega ^{2}\text{.}
\label{min2}
\end{equation}%
The fact that the frame (\ref{min}) gives us an example of the synchronous
frame is quite obvious if the nature of coordinates under the horizon is
taken into account. However, strange as it may seem, this simple fact was
not, to the best of our knowledge, pointed out in the textbooks on the
subject.

Thus, under the horizon we have two different synchronous frames (\ref%
{metlem}) and (\ref{min2}). The first one is the standard Lema\^{\i}tre one
extended across the horizon into the inner region. To obtain the second one,
we start from the metric (\ref{min}) already under the horizon with the help
of simple reparametrization of time, as explained above.

Now, we can return to our issue of maximizing the proper time using the
approach of the precedeng section. If we use frame (\ref{min}), the
singularity corresponds to $T=0=const$. It is clear that rescaling (\ref{tn}%
) does not change this fact, so also $\hat{t}=\hat{t}(0)$ =$const$ for the
singularity. Then, previous consideration applies directly with the
conclusion that the best choice is the geodesic with $y=const$, so $%
\varepsilon =0$. Meanwhile, the surface $\tilde{t}=const$ cannot coincide
with the singularity $r=T=0$. Indeed, eq. (\ref{tt}) shows that the surface $%
\tilde{t}=\tilde{t}_{0}$ represents some curve $T=T(y,\tilde{t}_{0})\,\ $or $%
y=y(T,\tilde{t}_{0})$ that does not coincide with $T=0$. It intersects the
singularity in one point where $y=y(0,\tilde{t}_{0})$.

The Lema\^{\i}tre frame is important in other class of problems which
involves both $T$ and $R$ regions, for example the description of the outer
world picture seen by an infalling observer (see Section VIII). From this
point we will consider only Lema\^{\i}tre synchronous frame in the present
paper, leaving the other synchronous frame (\ref{min}) for a separate study.

\section{River model and peculiar velocities}

For the description of particle kinematics, the concepts of river of space
and peculiar velocities against such a background are quite convenient. Our
approach is the counterpart of the "river model" \cite{ham}, see also eq.
(27) of \cite{zero}. Outside the horizon, the off-diagonal part of the
Painlev\'{e}-Gullstrand metric contains the quantity $v$ which can be called
"velocity of flow of space". More precisely, the rate with which $r$ changes
can be decomposed to two parts:%
\begin{equation}
\frac{dr}{d\tilde{t}}=-v+v_{p}\text{,}  \label{rtv}
\end{equation}%
where, by definition, $v_{p}$ is the peculiar "velocity" with respect to the
"flow". (To avoid misunderstanding, we would like to point out that, in
general, these velocities have nothing to do with those discussed in Sec. IV
of the present paper).

It is convenient to introduce the orthonormal tetrads in which the time-like
vector is directed along the velocity of the "flow"%
\begin{equation}
h_{(0)}^{\mu }=\frac{\partial }{\partial \tilde{t}}-v\frac{\partial }{%
\partial r}
\end{equation}%
and the other space-like vectors are directed along the coordinate axis:%
\begin{equation}
h_{(1)}^{\mu }=\frac{\partial }{\partial r}\text{, }h_{(2)}^{\mu }=\frac{1}{r%
}\frac{\partial }{\partial \theta }\text{, }h_{(3)}^{\mu }=\frac{1}{r\sin
\theta }\frac{\partial }{\partial \phi }.
\end{equation}%
Covariant components in coordinates ($\tilde{t},r,\theta ,\phi $) are equal
to%
\begin{equation}
h_{\mu (0)}=(-1,0,0,0),
\end{equation}%
\begin{equation}
h_{\mu (1)}=(v,-1,0,0)
\end{equation}

Let us consider pure radial motion, so $L=0$. We define tetrad components of
the velocity according to 
\begin{equation}
V^{(i)}=-\frac{h_{\mu }^{(i)}u^{\mu }}{h_{\mu (0)}u^{\mu }}\text{.}
\end{equation}%
Then, it is easy to obtain that $h_{\mu }^{(1)}u^{\mu }=v\frac{d\tilde{t}}{%
d\tau }+\frac{dr}{d\tau }$, $h_{\mu (0)}u^{\mu }=-\frac{d\tilde{t}}{d\tau }$%
, $V^{(1)}=v+\frac{dr}{d\tilde{t}}.$ Using (\ref{rtv}) we see that the
tetrad component of the velocity of a free moving particle just coincides
with the peculiar velocity:%
\begin{equation}
V^{(1)}=v_{p}\text{.}
\end{equation}%
\ This means that $v_{p}$ is not just a coordinate velocity, but is a
physical 3-velocity of a particle with respect to the Painlev\'{e}%
-Gullstrand frame. In particular, $v_{p}<1$.

In what follows, it is convenient to rewrite energy integral in terms of
free-fall and peculiar velocities. Taking into account (\ref{eu}) - (\ref{my}%
), (\ref{tt}) we have the following expression for the conserved momentum in
the GP frame 
\begin{equation}
\varepsilon =(1-v^{2})\frac{d\tilde{t}}{d\tau }-v\frac{dr}{d\tau }.
\end{equation}

Using the definition of $v_{p}$ this expression can be rewritten as 
\begin{equation}
\varepsilon =\frac{d\tilde{t}}{d\tau }(1-vv_{p}).  \label{etv}
\end{equation}

In the next section we will consider special properties of the trajectory
with $\varepsilon =0$. Here we can see that for such a trajectory $v_{p}=1/v$%
, and it exists only below the event horizon where $v$ is superluminal. It
is also easy to see that a positive $\varepsilon $ corresponds to $v_{p}<1/v$
and negative $\varepsilon $ requires $v_{p}>1/v$.

The first factor in (\ref{etv}) represents the Lorentz time dilation of an
object moving with respect to GP frame, so 
\begin{equation}
\frac{d\tilde{t}}{d\tau }=\frac{1}{\sqrt{1-v_{p}^{2}}}\text{.}  \label{ttv}
\end{equation}%
The validity of eq. (\ref{ttv}) can be checked directly if one uses eqs. (%
\ref{tz}), (\ref{rtv}) and (\ref{etv}).

Then, eq. (\ref{etv}) may be presented as 
\begin{equation}
\varepsilon =\frac{1-vv_{p}}{\sqrt{1-v_{p}^{2}}}.  \label{evv}
\end{equation}%
Taking derivative with respect to $v_{p},$ it is possible to get 
\begin{equation}
(1-v_{p})^{3/2}\frac{\partial \varepsilon }{\partial v_{p}}=v_{p}-v.
\label{v-v}
\end{equation}

As $v_{p}<1$ always, and $v>1$ inside the horizon, the right hand side of
this equation is negative under the horizon. Therefore, the sign of the
derivative is negative, so inside a horizon bigger peculiar velocity always
means lower value of $\varepsilon $ and vice versa. We will use these
results in what follows.

\section{Behavior of the Lema\^{\i}tre time}

Let us return to the Schwarzschild metric or its generalization (\ref{met}).
We are going to discuss the behavior of the Lema\^{\i}tre time during a fall
and the possibility of maximizing it. To the best of our knowledge, such
discussion was absent from literature before, and was only briefly mentioned
in \cite{austr}. We start this section with a comment about the difference
between the Lema\^{\i}tre time and the proper one. Sometimes in textbooks
the description of a free fall into a black hole both times are used as
synonyms, though in a general case this is not correct. It follows from eq. (%
\ref{metlem}) that the proper time of a particle free falling into a black
hole coincides with the Lema\^{\i}tre time if the value of $\rho $ is
constant for a given particle. This means that the particle falls with zero
velocity at infinity and has $\varepsilon =1$. This is also seen from eq. (%
\ref{evv}) with $v_{p}=0$. All particles with other values of energy have
their own proper times different from the Lema\^{\i}tre time. In particular,
it is so for particles with $\varepsilon =0$.

In contrast to the proper time, which is individual for each particle, the
Lema\^{\i}tre time is one of coordinates of the metric and can be considered
as a \textquotedblleft universal time\textquotedblright\ appropriate for
describing the free fall of any particle. Suppose we have an infinite chain
of observers with different $\rho $ unbounded from above, free falling in
such a way that their individual $\rho $ remain constant. As it is noted
above, all such observers measure the same proper time. As the proper time
for them is the Lema\^{\i}tre one, we can say that this chain measures the
Lema\^{\i}tre time. Note that a time scale for each individual observer is
finite. For example, in the Schwarzschild space time an observer with $\rho
=\rho _{1}$ hits the singularity $r=0$ at $\tau =\rho _{1}$ (if the constant
in the definition of time is chosen so that $\frac{3}{2}(\rho -\tau )=r_{+}$
on the horizon - see eq. 102.2 in \cite{LL}) and does not exist anymore.
However, for any Lema\^{\i}tre time there exist observers with big $\rho $
enough which can continue to measure it.

Now we turn to the question of the Lema\^{\i}tre time for the trajectory
with $\varepsilon =0$. This trajectory maximizes the proper time of a free
fall. Note, however, that getting $\varepsilon =0$ directly at the horizon
is impossible (it requires the peculiar velocity equal to the speed of
light), so this limit can be reached only asymptotically. For the Lema\^{\i}%
tre time we get even more interesting result. Using notations of the
previous section and the Eq.(21) we can write in general

\begin{equation}
\tilde{t}=\int_{0}^{r_1}\frac{dr}{v-v_{p}}\text{.}  \label{tint}
\end{equation}%
for the time needed to reach the singularity from $r=r_1$.

Using that $v_{p}=1/v$ at $\varepsilon =0$, we can easily rewrite this
integral in terms of the flow velocity $v$ only. This integral in the
Schwarzschild space time gives us

\begin{equation}
\tilde{t}=r_{+}\ln (\frac{{\sqrt{r_{+}}+\sqrt{r_1}}}{{\sqrt{r_{+}}-\sqrt{r_1}%
}})-2\sqrt{r_{+}r_1}.  \label{Ltime}
\end{equation}

Clearly, this integral diverges at $r=r_{+}$, so that the Lema\^{\i}tre time
for this particular trajectory (passed by the finite interval $(\pi /2)r_{+}$%
of the proper time) is infinite. Moreover, since bigger $v_{p}$ makes the
integral bigger, the Lema\^{\i}tre time for trajectory with $\varepsilon <0$
diverges as well.

That this time diverges is the consequence of the incompleteness of the Lema%
\^{\i}tre frame. It covers only the right $R$ region and the $T$ one. In a
similar way, another Lema\^{\i}tre frame can be built that covers the left $%
R $ region and the same $T$ one. This problem does not arise if one uses,
say, the Kruskal frame that is complete and covers all space-time (see, e.g. 
\cite{mtw}).

As far as the Lema\^{\i}tre time for any trajectory with $\varepsilon >0$ is
concerned, it is evidently finite, since the function under the integral is
finite at any point from the horizon to singularity.

\section{What can a falling observer see?}

The material of the previous sections gives us a new look at the old
question about possibility of an observer free falling into a black hole to
see the entire future of the Universe. This question is usually considered
as an example of a wide-spread mistake. Naively, since finite proper time $%
\tau $ can correspond to arbitrary large remote observer time $t$, we could
expect that the answer is "yes". However, as it is pointed out in
methodological literature previously, this argument fails if we take into
account the finiteness of speed of light. Then, the answer (at least for the
Schwarzschild black hole, where there are no inner horizons) becomes "no",
as is explained in \cite{kras}, \cite{gpu}.

\begin{figure}[h]
\centering
{\includegraphics[width=0.8\textwidth]{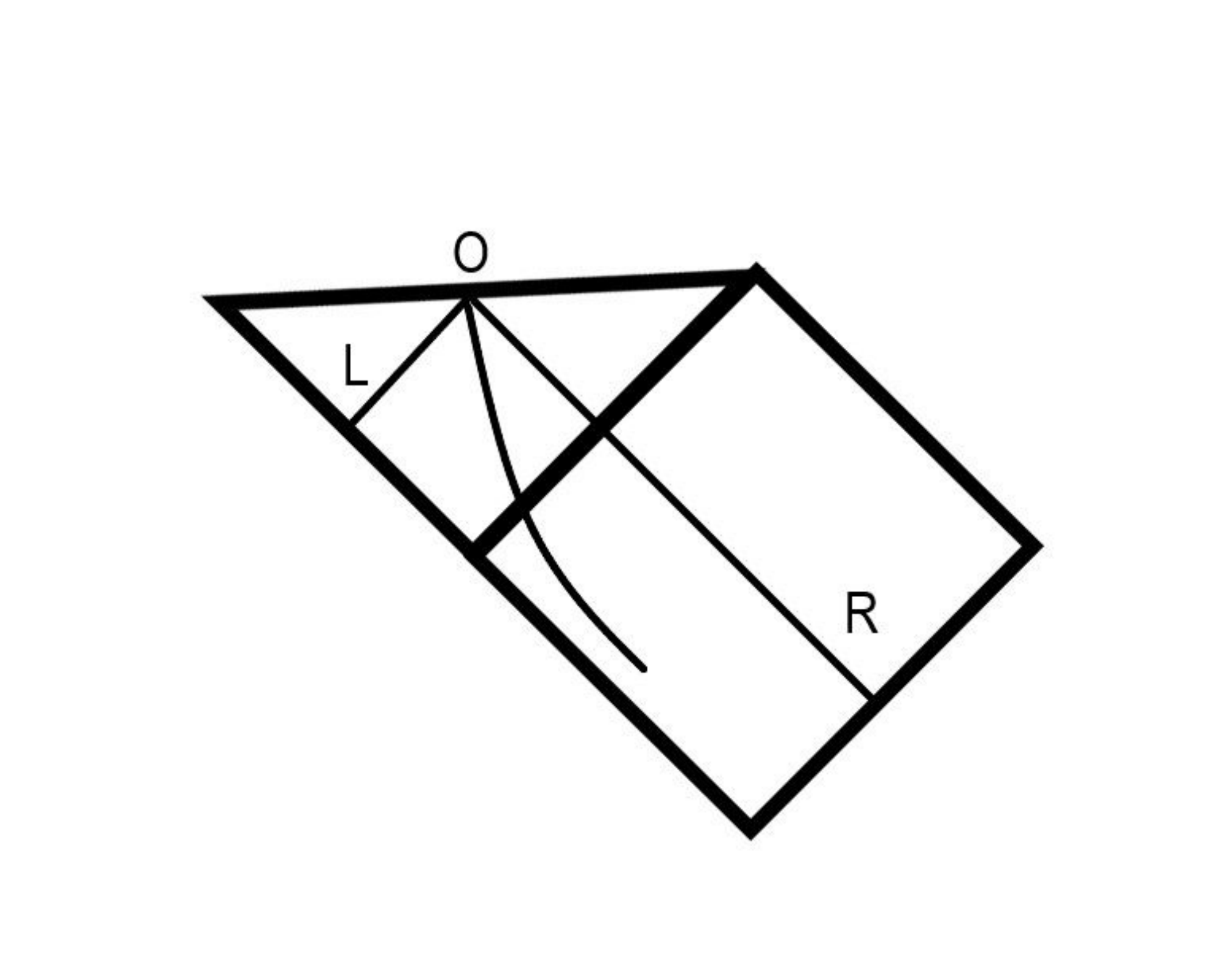}}
\par
\caption{ Trajectory of an observer inside a black hole and his individual
horizon.}
\end{figure}

However, this is not the end of story. There were additional assumptions
used in the above references without reservations: (i) an observer falls in
a black hole starting from the $R1$ region, but not from the $R2$ region
(this is a natural assumption for a realistic observer) and (ii) he/she
moves along the geodesic path. If one of these requirements (or both) is
violated, the answer is to be modified (see below). This is illustrated in
Fig.1 showing an observer falling into a singularity and some of
light-cones. The part of space-time which can be seen by the observer lies
within the past light cone with the vertex at the point $O$ where he/she
hits the singularity at some time $t_{s}$. Any events outside this light
cone cannot be seen by the observer. So, we can treat this cone as an
individual event horizon for the observer in question. Note, that this
construction is well known in cosmology, and corresponds to the event
horizon of an observer in contracting Universe -- it exists due to the fact
that (i) lifetime of an observer before the end in cosmological singularity
is finite, (ii) the singularity is space-like, so, the range of events which
can be seen is finite too. In black hole physics is it important not to mix
up this individual horizon (which can be used in description of a free-fall)
with standard observer-independent event horizon $r=r_{+}$ absent in
cosmology.

In our two-dimension diagram the cross-section of this light cone is in fact
two light geodesics directed towards the past. They are represented by the
left and right rays in the diagram, so the observer can see events inside
this cone at the time of singularity hitting. In the simplest case of the
Schwarzschild metric we can explicitly integrate the equation for light
geodesics and get the analytical form of the light rays in question. Indeed,
integration of equation $ds^{2}=0$ with (\ref{gp}) taken into account gives
us

\begin{equation}
\tilde{t}_{R}=\tilde{t}_{s}-r+2\sqrt{r_{+}r}-2r_{+}\log \frac{(\sqrt{r}+%
\sqrt{r_{+}})}{\sqrt{r_{+}}}\text{,}  \label{1+v}
\end{equation}%
\begin{equation}
\tilde{t}_{L}=\tilde{t}_{s}-r+2\sqrt{r_{+}r}+2r_{+}\log \frac{(\sqrt{r_{+}}-%
\sqrt{r})}{\sqrt{r_{+}}}.  \label{1-v}
\end{equation}%
Here, subscripts "L" and "R" refer to the left and right rays, respectively.
The constant of integration is chosen to ensure that $\tilde{t}=\tilde{t}%
_{s} $ at $r=0$.

The zone of events which an observer can see before crash into the
singularity is given by the interior of the corresponding light cone. For a
given $r$, it is given by inequality $\tilde{t}_{L}\leq \tilde t\leq \tilde{t%
}_{R}$. Clearly, for a fixed $r$ we have $\tilde{t}_{R} \to \infty$ if $%
\tilde {t}_{s} \to \infty$

At this point we remind a reader that any massive particle moving from $%
R_{1} $ zone into $T$ zone along a geodesic has $\varepsilon >0$. Now, let
an observer come from $R_{1}$, being equipped with some engine. Let him turn
it on in the point $r$ under the horizon in the $T$ region in such a way
that he sits on the trajectory $\varepsilon =0.$ He turns the engine off
afterwards, thus making a transition from the geodesic trajectory with $%
\varepsilon >0$ to the one with $\varepsilon =0$. Then, the Lema\^{\i}tre
time from $r$ to singularity is given by (\ref{Ltime}). 

Let $r\rightarrow r_{+}$. Then, this time diverges in accordance with (\ref%
{1-v}). In the point where a particle (astronaut)\ hits a singularity, we
have in (\ref{1+v}), (\ref{1-v}) $\tilde{t}_{s}\rightarrow \infty $. In
doing so, the trajectory of an astronaut bends more and more to the horizon,
when $r$ is taken to be closer and closer to $r_{+}$. And, the part of the
external Universe in the $R_{1}$ region becomes more and more visible (see
Fig.2)

\begin{figure}[h]
\centering
{\includegraphics[width=0.8\textwidth]{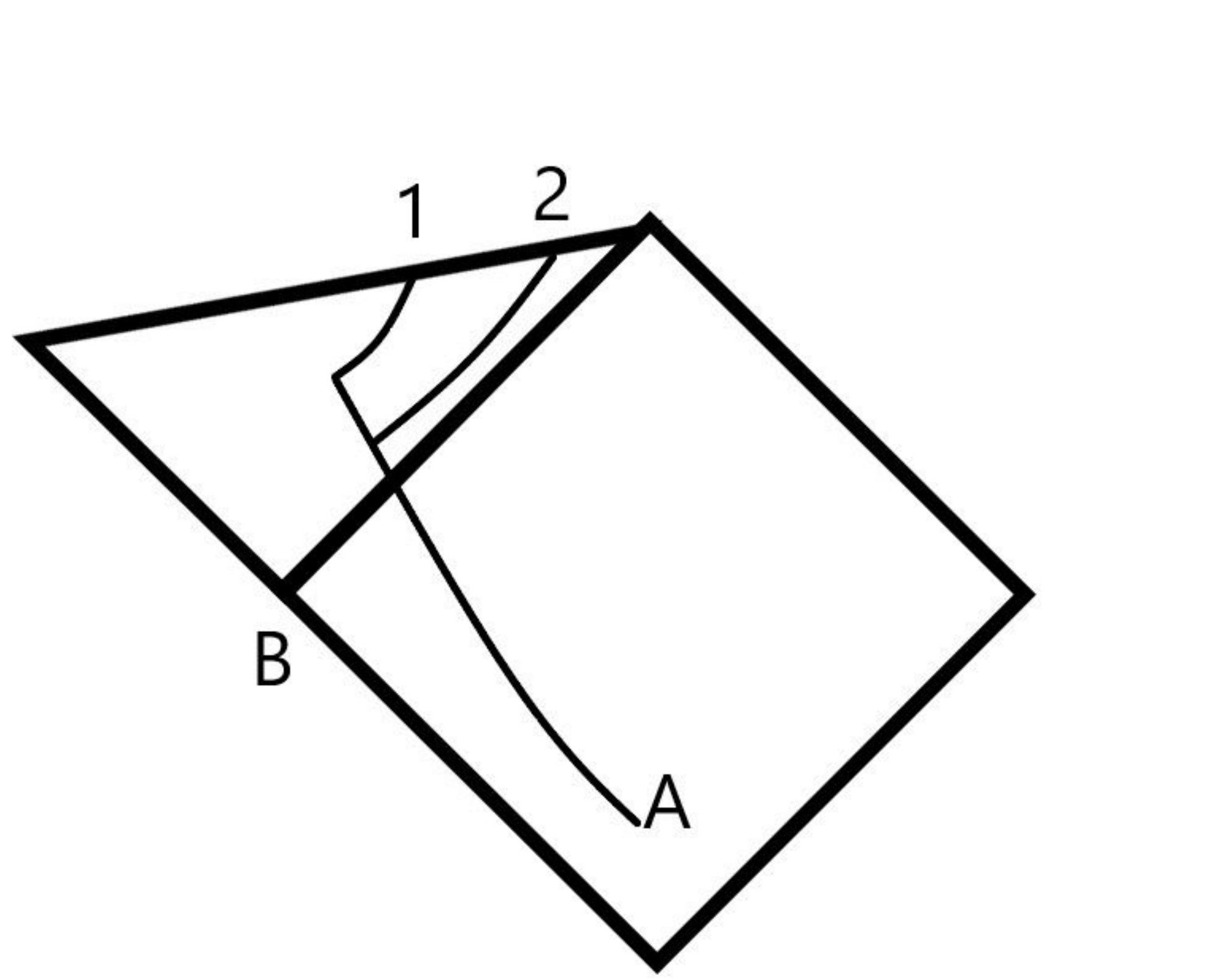}}
\par
\caption{Schematic picture of trajectories of an astronaut changing his path
to the E=0 trajectory. }
\end{figure}

Thus, although the future history of the Universe, accessible to an observer
for each astronaut with a rocket, is finite, this part is not limited by
some fundamental restrictions. It is limited by the power of the engine
required to change an initial trajectory and sit on the trajectory $%
\varepsilon =0$. It is also limited by the ability of the observer to
survive under enormous acceleration caused by this engine: the more powerful
is the engine (provided the observer survives during its work), the bigger
slope of the $\varepsilon =0$ to the horizon be, and the bigger part of the
future of the Universe the observer will see. In other words, each concrete
observer sees only a finite part of the future but this part can be expanded
without restrictions. One can call this situation "an observer sees the
whole future of the Universe in a weak sense".

In the limit under discussion the trajectory $\varepsilon =0$ would coincide
with the horizon but the only physical object which can reach it there is a
light ray (or any another beam of massless particle). Since both velocities $%
v_{p}=1$ and $v=1$ (as is explained in this Section above), it is seen from (%
\ref{rtv}) that the coordinate $T$ is a constant and so is the radial
coordinate $r=r_{+}$ - the beam stays at the horizon forever! However, a
realistic photon emitted exactly at the horizon will experience redshift
growing exponentially with the Lema\^{\i}tre time \cite{along}, so the
validity of the geometric optics approach becomes questionable and requires
separate treatment.

Consider again the view seen by an infalling astronaut. The singularity lies
in absolute future, so an astronaut cannot see it. The right ray OR comes
from the outer part of Universe from which the astronaut arrived. The
astronaut can see it in point O only where this ray hits the singularity.
Meanwhile, the ray OL comes from the $R2$ region ("mirror world"). It is
inaccessible to the outer observer in region $R1$ but an astronaut in the $T$
region, i.e. inside a black hole, can see it. If there exists the left $R$
region ($R2$) from which information is allowed to propagate, an astronaut
can see corresponding events. If there is no such a region (say, in the case
of collapse), an astronaut, at least, can see light emitted by the surface
of the collapsing star ahead of him (see Ref. \cite{Hamilton} for details).
He can also see the right horizon (which is called a true horizon in \cite%
{Hamilton}) looking in the backward direction, the redshift being finite.

\section{Two strategies of an astronaut}

The unboundedness of the Lema\^{\i}tre time in free fall into a black hole
has one more interesting feature, if we consider not a view of the "outer"
world for an observer, but ask the question is it possible for the astronaut
under discussion to communicate with it. Evidently, a communication with any
object staying permanently outside the horizon becomes impossible once the
astronaut crosses horizon (in fact, even earlier). However, what about
objects falling into the black hole? We do not consider here this question
in detail, but illustrate with the simplest example of an object with
constant spatial Lema\^{\i}tre coordinate $\rho $. A particle with $\rho
=\rho _{i}$ reaches the singularity at $\tilde{t}_{1}=\rho _{i}$. If $\tilde{%
t}_{1}<\tilde{t}_{s}$, the particle in question will pass through the
astronaut on the way to singularity. So that, if $\tilde{t}_{s}$ is very
big, particles with rather big $\rho $ can still pass near the astronaut in
question and could literally touch him/her by hand. Since two both
participants (an astronaut and a particle) meet at the same point, the
possibility of mutual communication is evident at least till this point
which occurs somewhere under the horizon. So that, if $\tilde{t}_{s}$ is
very large, more and more free falling observers with bigger and bigger $%
\rho =const$ are able not only to see the astronaut but mutually communicate!

So, if one uses an engine near the horizon to make $\varepsilon =0$, this
helps in achieving two goals at once: maximizing proper time till the
singularity and maximizing the possible future of the Universe seen during
this fatal fall. We should remark, however, that if the horizon is already
passed, and the observer is in $T$ region, these two goals may require
different strategies. For example, suppose that the observer inside the
horizon found himself at a trajectory with $\varepsilon =0$, but some fuel
remains. Is it reasonable to use the fuel more? If we want to make the
proper time before hitting singularity as large as possible, the answer is
obviously "no" -- the trajectory with $\varepsilon =0$ is optimal. But what
about the Lema\^{\i}tre time till the singularity?

The time from some initial $r=r_{0}$ to the singularity $r=0$ is given by (%
\ref{tint}) with $r=r_{0}$. Under the horizon the integrand is positive, $%
v>v_{p}$ (see discussion after eq. (\ref{v-v})). Therefore, the bigger $%
v_{p},$ the bigger is the Lema\^{\i}tre time. So, the astronaut should use
the remaining fuel -- the fight against gravity makes sense! Ironically, not
for the fighter -- his proper time till singularity decreases while Lema%
\^{\i}tre time increases.

Using the results below Eq.(\ref{v-v}) we may also note that if an astronaut
understands that he/she is actually on the trajectory with $\varepsilon <0$
and wants to achieve the maximum possible proper time, it is necessary to
decrease $v_{p}$ in order to reach $\varepsilon =0$. On the contrary, such
an astronaut should increase $v_{p}$ as much as possible to maximize the Lema%
\^{\i}tre time (allowing to see more future of the outer word).

In other words, a researcher inside the horizon should pay by the time of
his own life for satisfying his curiosity!

\section{Conclusions}

In our paper we have considered two strategies for an observer falling into
a spherically symmetric black hole. The observer may try to maximize either
the proper time (this is the case usually considered in textbooks) or the
Lema\^{\i}tre time. The latter goal allows the observer to see as much
future history of the Universe as possible. These two different goals may or
may not lead to different strategies. Of course, they coincide for an
observer in the outer region --- switching on the engine the observer could
( if the engine is powerful enough) avoid entering inside the horizon. Then,
both the proper and Lema\^{\i}tre lifetimes tend to infinity. At the horizon
these two goals still require the same action --- use the engine us much as
possible getting the biggest possible positive peculiar velocity. In the
unreachable limit this strategy would turn the energy of the observer to
zero, maximizing the proper time to the singularity. Simultaneously, this
would lead to peculiar velocity of the observer equal to the speed of light
and the Lema\^{\i}tre time to singularity equal to infinity. So, an
astronaut with powerful engine which succeeds in ( almost) maximizing the
proper lifetime will see as, a \textquotedblleft bonus\textquotedblright ,\
(almost) complete future history of the outer world.

Let, being already inside the horizon, the observer decide to live as much
as possible and, thus, to maximize the proper time till the singularity.
Then, the best strategy, according to (\ref{tau}) consists in reaching the
zero energy and turning the engine off afterwards. If, however, the observer
would like to see as much future of the outer world as possible, the
strategy is to maximize the peculiar velocity (see equation (\ref{tint})).
This means that such a curious observer should use all the power of the
engine and not switch it off deliberately. The closer $v_{p}$ is to the
unreachable speed of light, the bigger part of future such an observer can
see. Such an observer increasing his peculiar velocity as much as possible
under the horizon would have large negative $\varepsilon $ (see formula (\ref%
{evv}), where the numerator is finite and negative for this case, while the
denominator is very close to zero for $v_{p}\approx 1$). Then, it follows
from (\ref{tau}) with very big $\left\vert \varepsilon \right\vert $ that
this remote future will pass almost instantly by observer's own clock
between the initial point and the singularity. We would like to stress that
there is intimate connection between some properties of motion and metric.
Namely, when a point where an astronaut begins to follow the line with $%
\varepsilon =0$ approaches the horizon, the Lema\^{\i}tre time diverges and
the part of the outer universe available for an astronaut increases.

In the present methodological paper we considered an idealized situation and
did not take into account details presented in any realistic scenarios. For
example, any realistic engine has a finite power, so reaching the trajectory
with $\varepsilon =0$ exactly at horizon is impossible. When precisely it is
better to switch the engine on, requires a special calculations. More
realistic model should also include the fact that an engine works not
instantly, but rather gives a finite acceleration. We leave such more
technical questions to a separate investigation.

Our consideration applies to any spherically symmetric metric with a simple
(nonextremal)\ horizon, which is static outside it and homogeneous inside,
and to any metric theory of gravitation provided particles in the absence of
external forces move along geodesics.

\begin{acknowledgments}
The work was supported by the Russian Government Program of Competitive
Growth of Kazan Federal University
\end{acknowledgments}

\end{document}